**Exploring 2D Materials by High Pressure Synthesis: hBN, Mg-hBN, b-P, b-AsP, and GeAs**

N.D. Zhigadlo

*CrystMat Company, CH-8037 Zurich, Switzerland*

**Abstract**

In materials science, selecting the right synthesis technique for specific compounds is one of the most important steps. High-pressure conditions have a significant effect on the crystal growth processes, leading to the creation of unique structures and properties that usually are not possible under normal conditions. The prime objective of this article is to illustrate the benefits of using high-pressure, high-temperature (HPHT) technique when developing two-dimensional (2D) materials. We could successfully grow bulk single crystals of hexagonal boron nitride (hBN) and magnesium doped hexagonal boron nitride (Mg-hBN) from Mg-B-N solvent. Further exploration of the Mg-B-N system could lead to the crystallization of isotopically $^{10}$B and $^{11}$B enriched hBN crystals, and other doped variants of it. Black phosphorus (b-P) and black phosphorus doped with arsenic (b-AsP) were obtained by directly converting its elements into melt and subsequently crystallizing them under HPHT. Germanium arsenide (GeAs) bulk single crystals were also obtained from the melt at a pressure of 1 GPa. Upon crystallization, all these compounds exhibit the anticipated layered structures, which makes them easy to exfoliate into 2D flakes, thus providing opportunities to modify their electrical behavior and create new useful devices.

# 1. Introduction

Growing two-dimensional (2D) materials is a challenging task that involves careful control of the growth conditions and requires specific techniques and methods. Different 2D synthesis techniques have varying benefits and drawbacks, such as synthesis time constrains, product purity, and potential for large-scale production [1-3]. Below we outlined a few difficulties and problems that arise with the growth of 2D materials: (1) During the growth process, unwanted contaminants or foreign substances might easily infect the 2D material [4]. Since these impurities might impact the material's characteristics and performance, maintaining a clean growth environment is crucial. (2) It is difficult to achieve controlled nucleation and growth of 2D materials. To obtain 2D materials of the desired quality, it is necessary to adjust parameters like temperature, pressure, precursor concentration, and growth time [5]. (3) A lot of 2D materials need specific substrates for growth. In order to produce 2D materials of the highest quality, it is essential to select a substrate with the appropriate lattice match and surface characteristics [6]. (4) While small-scale growth of 2D materials is feasible in laboratories, scaling up the process for large-scale production remains difficult [3]. Methods that work effectively on a small scale may not be easily adaptable to industrial-scale manufacturing. (5) Achieving a uniform growth of 2D materials across large areas is crucial for practical applications [7]. To ensure consistent material properties and performance, the growing process must be reproducible. (6) Some methods of growth involve the use of poisonous or dangerous chemicals [8]. When working with such materials, it is essential to follow safety procedures and limit any harmful effects on people and the environment. (7) Not all 2D materials have well-established growth strategies, so different 2D materials may need unique growth procedures. Exploring novel growth approaches for less studied 2D materials is a never-ending task [1]. (8) Some 2D materials may degrade when exposed to air, either through oxidation or by interaction with moisture [9]. Maintaining the stability of these materials during growth and subsequent handling is crucial for their use in practical applications.

Despite all these difficulties, considerable progress has been achieved in the growth of various 2D materials. Thanks to advancements in growth techniques and to our understanding of the underlying mechanisms, the wide range of 2D materials that are currently available is bound to increase. The readers are invited to read the review papers describing the most recent progress in this field [1,2,10-12].

The growth of 2D crystals under high pressure is an exciting research area of materials science and condensed-matter physics, which is less explored compared to other growth methods [1,2]. Below we highlight some key aspects related to the growth of 2D materials under high pressure [13,14]: (1) Pressure-assisted exfoliation: Exfoliating layered materials, such as graphene from graphite or transition-metal dichalcogenides from their bulk counterparts, is a typical technique for producing 2D flakes. High pressure can facilitate the exfoliation process, making it easier to split the bulk crystal into thin layers. (2) Under high pressure, some materials, like phosphorus for example, may undergo phase transitions or structural changes that are not possible at ambient conditions. These new structures may lead to the discovery of novel 2D materials with distinctive electrical, optical, and mechanical properties. (3) High pressure can enhance the alignment of atomic layers and improve the crystallinity and quality of 2D materials. The increased pressure provides a more controlled environment for the crystal growth, reducing defects and disorder in the resulting 2D structures. (4) The band structure and electronic properties of 2D materials can be significantly altered under high pressure. This opens up possibilities for modifying their electronic behavior and creating new functional devices. (5) Application potential: Due to their unique properties and structures, 2D materials grown under high pressure may be used in a variety of industries, including electronics, photonics, sensors, and energy storage [13,14].

Although high pressure may look like an expensive procedure, due to the low energy conveyed by pressure, the actual operational costs are comparable to those of other solid state synthesis methods. For example, the power consumption of our cubic anvil apparatus is influenced by factors such as the pressure and temperature, the volume of the sample being processed, the length of the experiment, and the efficiency of the apparatus itself [15]. In each of our growth experiments the power consumption was less than 1.5 kW and the growth process was relatively short. These conditions are beneficial for energy-efficient operation. Therefore, the application of the HPHT method to the mass production of 2D materials cannot be underestimated. One of the best examples in this direction is the commercial production of synthetic diamonds since some decades [16].

In this paper, we report on the successful growth of hBN, Mg-doped hBN, black P, As-doped black P, and GeAs bulk crystals under HPHT conditions. Our reasoning for studying those compounds stems from the facts that, while they show great potential for various practical applications, questions about their synthesis and purity still need to be explored. Some of those existing problems are briefly discussed in the results and discussion section. The obtained results provide valuable experimental insight about the high-pressure growth mechanisms.

## 2. Experimental details

For the synthesis of 2D materials described here we used a semi-cylindrical multi-anvil module manufactured by Rockland Research Corporation (USA). It consists of eight large outer steel dies squeezing six small inner tungsten carbide (WC) anvils, thus applying an amplified pressure on the innermost cubic cell. A general overview of the cubic-anvil apparatus and a cross-sectional view of the high-pressure sample-cell assembly used in the high-pressure growth is schematically shown in Fig. 1.

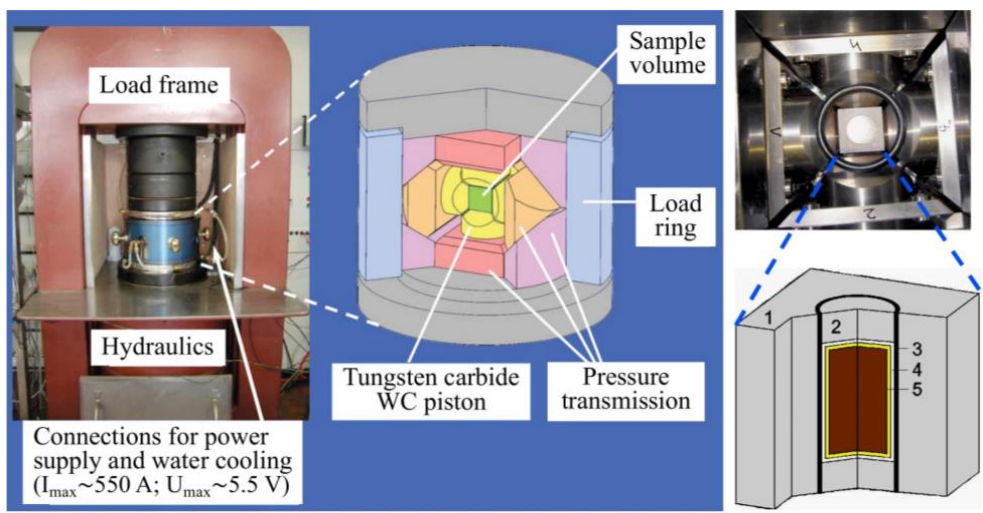

**Figure 1.** Cubic anvil type apparatus and high-pressure cell assembly: 1,2-pyrophyllite, 3-graphite heater, 4-boron nitride (BN) crucible, 5-sample. The pyrophyllite cube is placed in the core of the cell between the tungsten-carbide (WC) anvils. The power supply source permits to gain up to ~550 amps of current and ~5.5 volts of voltage, which corresponds to the power consumption of about 3 kW.

The sample-cell assembly has a tubular graphite heater placed in the bore of a pyrophyllite cube. The heater is loaded with a cylindrical BN crucible in the center and with pyrophyllite tablets on top and bottom. The graphite heater is heated by a current that flows through it, provided from outside through the steel parts and the two WC anvils. A water-cooling system prevents the WC anvils from overheating. More information regarding the high-pressure apparatus and the experimental setup can be found elsewhere [17, 18]. The high-pressure method was successfully used earlier by us to grow crystals of a wide range of superconducting- [19-22] and magnetic materials [23,24], diamonds [25], cuprate oxides [26-29], pyrochlores [30,31], *Ln*Fe*Pn*O (*Ln*: lanthanide, *Pn*: pnictogen) oxypnictides [32-37], polymers [38], and a number of other compounds [39-42].

High purity elements of red P (99.999%), grey As (99.999%), Ge (99.999%), Mg flakes (99.99%), amorphous B (99.99%) and BN (99.99%) were used as starting reactants. Commercial amorphous B, and BN products, including BN crucibles were purified by firing them in a radio frequency induction furnace for several hours at ~2000 °C under dynamic vacuum conditions to remove boron derivatives and other volatile contaminants. In this process, an alternating current is passed through a coil to heat a tungsten crucible that has been filled with materials. For each compound the starting reactants were thoroughly mixed in the glove box and pressed into pellets of about 8 mm in diameter and 10 mm in length. The pellets were then placed in a cylindrical BN crucible surrounded by a graphite-sleeve resistance heater and inserted into a pyrophyllite cube, which served as a pressure transmitting medium. The pyrophyllite cell was then placed inside the high-pressure setup as depicted in Figure 1 and was cold-pressurized to 1-3.5 GPa. The optimum growth conditions provided in the following section were adjusted for each individual composition by varying the heating temperature, reaction time, applied pressure, and cooling rate. The conversion rate for b-P crystals was found to be the highest, reaching about 80%, whereas it was lower for other materials. Following the completion of the crystal growth procedure, the produced products were mechanically removed, and their characterization was carried out using a variety of common laboratory instruments.

The morphology and dimensions of the produced crystals were evaluated using a Leica M 205 C optical microscope. The chemical composition of the crystals was determined using energy-dispersive X-ray spectroscopy (EDS, Hitachi S-3000 N), by averaging five measurements for each crystal. X-ray measurements of large-size crystals were performed in the Bragg-Brentano geometry at room temperature with a STOE StadiP diffractometer and Cu K$_\alpha$ radiation ($\lambda$ = 1.5406 Å). A micro-Raman spectrometer equipped with an argon ion laser with a wavelength of 514.5 nm was used to perform Raman spectroscopy measurements.

*Caution*: When working with arsenic-containing products outside the glove box, appropriate personal protection equipment should be used, since arsenic can form arsine gas, which is poisonous. When the synthesis of As-based compounds is complete, crucibles must be opened in a well-ventilated fume hood to prevent any potential contact with the volatile gas.

## 3. Results and discussion

*3.1. Crystal growth and characterization of hBN and Mg-doped hBN*

In recent years, with the advent of graphene, hBN has proven to be an excellent gate and substrate, offering the smooth and flat surface necessary for maximizing carrier mobility in graphene devices [43,44]. In addition to a large range of technical applications, the combination of excellent properties of hBN, together with the ability to assemble it in more complex artificially stacked structures, opens a new paradigm in the physics of 2D materials [45].

So far, the most successful method to produce single crystals of hBN is the HPHT growth by using different kinds of solvents [46], although some promising results were obtained also at ambient pressure conditions [47,48]. Transition metals like Fe, Ni, and Co, as well as their alloys with other metals, are frequently utilized as metallic solvents for the synthesis of hBN crystals [46-48]. At high pressure synthesis these solvents often have high thermal conductivities, which helps in dissipating heat produced during the HPHT process. This is crucial to maintain a uniform temperature distribution within the reaction crucible and to prevent local overheating, which could result in unwanted reactions or crystal defects. Metallic solvents are chemically inert under HPHT conditions, they usually do not react with the boron nitride precursor materials or the resulting hBN crystals, minimizing impurities and ensuring high crystal quality [47,48]. At HPHT conditions metallic solvents may enhance the solubility and reactivity of boron nitride precursor materials, making it easier to achieve the required crystal growth conditions and morphology. They can act as catalysts, promoting the synthesis of hBN crystals by facilitating the formation of boron nitride bonds or enhancing crystal nucleation and growth processes. This can lead to faster reaction rates and higher yields of high-quality hBN crystals.

In 1960s, HPHT was initially proposed by Wentorf [49] for the synthesis of cubic boron nitride (cBN). Later, since 1990s, Taniguchi et al. [50] used this method to produce cBN at 7.7 GPa and 2100 °C without any sintering aids. One of the first attempts to grow hBN crystals was carried out by Ishii and Sato [51], who used boron dissolved in a silicon flux under nitrogen atmosphere. However, due to carbon impurities, these hBN crystals were yellow in appearance and contained nitrogen vacancies. Later on, successful results in growing hBN crystals at atmospheric pressures were achieved by Kubota et al. [52,53], by using various kinds of solvents (Ni, Ni-Mo, and Ni-Cr). In 2004, hBN crystals were prepared by using purified Ba-B-N solvent system [54]. However, even the best hBN grown crystals often contain impurity regions, which can be identified by far-UV fluorescence microscopy [55]. Such regions most likely form due to residual oxygen and carbon contamination in the solvent and the starting ingredients. In practice, work with such systems requires specific environmental conditions. Therefore, the current challenge in the field is the search for alternative solvents which can still yield high-quality hBN crystals.

As an alternative we employed Mg-B-N as a solvent for growing hBN crystals. The distinct growth mechanism involves the formation of an intermediate magnesium nitride phase ($MgNB_9$) [56]. A mixture of Mg flakes, B and BN powders in a molar ratio 1:1.2:0.1 were used as starting materials. They were thoroughly mixed and pressed into a pellet before being placed into a BN crucible. The typical growth process entails the following steps: (i) compressing sample to 3 GPa at room temperature; (ii) increasing the temperature to 1950 °C in 1 h; (iii) dwelling at this temperature for 1-2 h; and (iv) lowering the temperature and pressure in 1 h. Once the crystal growth procedure is complete, the synthesized molten lump is typically heated for ~0.5 h to 750 °C in a quartz ampoule under dynamic vacuum to evaporate the excess of magnesium. As a result, $MgB_2$ crystals sticking together with BN crystals are produced (Figure 2(c,d)). The most suitable conditions for growing sizable bulk hBN crystals were found above 2.5 GPa, in the temperature range comprised between 1800 and 2100 °C, and the largest crystals occurred at ~1950 °C. Thus, by using the Mg-B-N system, hBN crystals can be produced at pressures as low as 2.5 to 3 GPa, compared to the 4 to 5.5 GPa range needed when using the Ba-B-N system [54]. An advantage of the lower pressure is the prevention of formation of mixed cBN and hBN phases, which occur when using the Ba-B-N solvent.

The effectiveness of the Mg-B-N can be seen in the large lateral dimensions of at least 2 mm or more of the hBN crystals that could be mechanically separated from the final product (Figure 2). Their high optical quality was verified by Koronski et al. [57]. It should be mentioned that flakes exfoliated from such bulk crystals were successfully utilized for the fabrication of van der Waals heterostructures with other layered materials [58].

The grown hBN bulk crystals exhibited strong, narrow peaks for the (002), (004) and (006), corresponding to well-ordered stacked crystal planes in the *c*-direction (Fig. 2(e)). The Raman spectrum of hBN crystal is shown in Fig. 2 (f). Only one Raman line with the full width at half maximum (FWHM) of about 8.2 cm$^{-1}$ at 1367.1 cm$^{-1}$ is observed. This peak corresponds to the $E_{2g}$ in-plane vibration mode of hBN. The FWHM value we measure is one of the narrowest that has been reported for hBN crystals [53], which suggests good crystal quality and homogeneity.

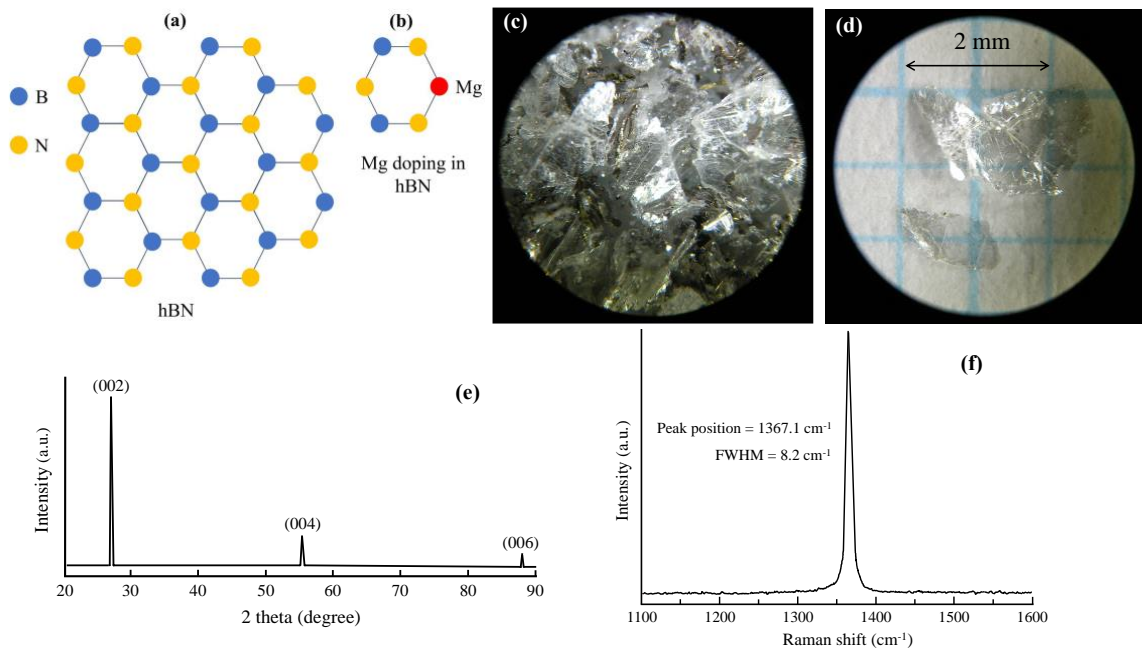

**Figure 2.** (**a**) Crystal structure of hBN and (**b**) Mg-doped hBN. (**c**) An optical image of the top-down perspective of the high-pressure products after removing the excess of Mg through vacuum annealing. Two main crystalline phases are visible: transparent and colorless hBN single crystals and black and gold-colored MgB$_2$ single crystals. (**d**) Image of hBN crystals extracted mechanically from the high-pressure products. (**e**) X-ray diffraction pattern of hBN bulk single crystal. (**f**) Raman spectrum of hBN bulk single crystal. The peak at 1367.1 cm$^{-1}$ with FWHM of 8.2 cm$^{-1}$ is indicative of the high purity and order of the crystals grown by this method.

One of the beneficial points is that Mg-B-N is also suitable for growing $^{10}$B- and $^{11}$B- enriched hBN (i.e., h$^{10}$BN and h$^{11}$BN), since this solvent consists of a mixture of elemental Mg, elemental B, which can be natural, $^{10}$B, $^{11}$B, and BN. The latter has very little impact on the final composition, as only a tiny concentration is used (1Mg + 1.2B + 0.1BN) for the preparation of the solvent. Because the $^{10}$B-isotope has one of the highest neutron absorption cross sections, the $^{10}$B-enriched hBN has a great potential for neutron detection [59].

By manipulating the starting composition of Mg-B-N solvent it is also possible to grow of *p*-type hBN using Mg-atoms as substitutional impurities in the hBN honeycomb lattice. Recently [60], by using micro-Raman spectroscopy, angle-resolved photoemission measurements and Kelvin-probe force microscopy we demonstrate that Mg dopants can significantly alter the electronic properties of hBN by shifting the valence-band maximum by 0.15 eV toward higher binding energies with respect to pristine hBN. We further show that Mg-doped hBN exhibits a robust, almost unaltered, band structure compared to pristine hBN. Such stable *p*-type doping of large-band hBN is a key feature for 2D material applications in deep ultra-violet light-emitting diodes or wide bandgap optoelectronic devices. Our findings demonstrate that conventional semiconductor doping by Mg as a substitutional impurity is a promising route to high-quality *p*-type doped hBN films.

*3.2. Crystal growth and characterization of b-P and b-AsP*

b-P and b-AsP have great potential for uses in optoelectronic applications that benefit from their anisotropic character and the ability to tune their band gap as a function of the numbers of layers and As content [61].

While most of the b-P properties are independent of how it was made, Iwasaki et al. [62] found that a two-dimensional Anderson localization is present in the electrical characteristics of needle-shaped b-P crystals prepared by the Bi-flux method, but not in crystals prepared by the HPHT method. This finding suggests that the growing technique affects the characteristics of b-P at low temperatures. The majority of b-P and b-AsP crystals were produced by using the Bi, Sn or some other variations of metallic fluxes [63]. A frequent byproduct of such flux growth is the Bi- or Sn-based foreign phases that have an impact on the crystallization process. Occasionally elemental Sn can also incorporate as an impurity into the crystal structure of the b-P and b-AsP final products and thus significantly affect their physical properties. For example, the presumed superconductivity of intercalated b-P, turned out to be related with traces of remaining Sn [64,65].

In the Bi-flux method, red-P cannot be used, since it does not dissolve in liquid Bi. On the other hand, white-P with a high degree of purity is uncommon due to its chemical activity. Commercial white-P needs to be purified, and it is quite challenging to get rid of residual impurities like S, Se, and As. Furthermore, white-P should not be exposed to air because it is highly reactive, toxic, and combustible. Overall, the high vapor pressure of P and As complicates the chemistry of phosphides and arsenides at high temperatures. Precautions should be taken to prevent breathing the fumes released when the ampules are opened. An alternative to the Bi and Sn fluxes is to use salt fluxes, such as NaCl/KCl, CsCl, and $ZnCl_2$. Unfortunately, the elemental solubility in salts fluxes is not particularly high.

The key advantage of the HPHT method is the direct growth of b-P and b-AsP crystals from its elements without adding any flux. Shirotani et al. [66] was able to grow b-P crystals by converting red-P to b-P and subsequently crystallizing b-P from its melt under a HPHT. Here, we report the growth of b-P and the in-situ doping of b-AsP through direct synthesis from a mixture of r-P and As dopant by high-pressure synthesis (Figure 3). In our experiments, pellets containing high purity powders of r-P for the growth of b-P crystals and grey As and r-P with molar ratio 30:70 for the growth of b-AsP were enclosed in a BN container and placed into a graphite heater. A pressure of 1 GPa was applied at room temperature. Then, by maintaining the pressure constant, the temperature was raised to 1050 °C in 2 h, held there for 3 h, then decreased to 650 °C in 10 h, and finally quenched to room temperature by switching off the heater.

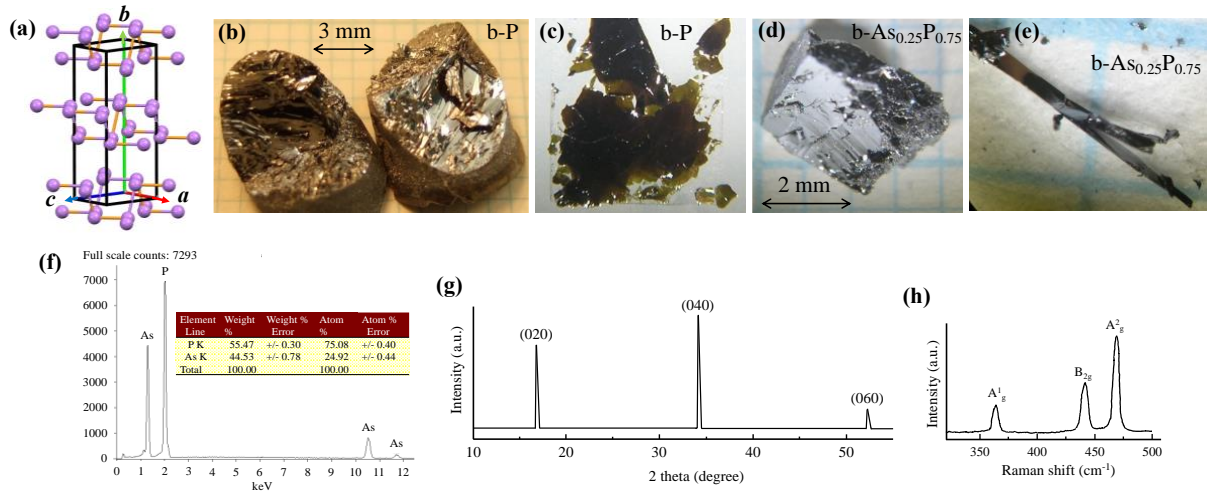

**Figure 3.** (**a**) Crystal structure of black phosphorus (b-P). (**b,c**) Optical images of high pressure grown bulk b-P lump broken into two pieces and b-P scotch exfoliated flakes. (**d,e**) Optical images of the bulk b-AsP crystalline lump and the exfoliated fragment of b-AsP. (**f**) Electron dispersive spectroscopy analysis of a b-As$_{0.25}$P$_{0.75}$ single crystal growth by HPHT method. The material contains only As and P with table indicating a composition that is close to As$_{0.25}$P$_{0.75}$. These values are the average of the data collected from five different spots on the crystal. (**g**) X-ray diffraction pattern of b-P bulk single crystal, which confirms its high crystallinity. (**h**) Raman spectrum of b-P bulk single crystal.

Single-crystal X-ray diffraction measurements confirmed the single-phase nature of the grown crystals. At HPHT conditions, b-P and b-AsP crystallize in an orthorhombic structure with *Cmca* space group and lattice parameters $a = 3.336(1)$ Å, $b = 10.210(5)$ Å, and $c = 4.393(2)$ Å for b-P and $a = 3.388(2)$ Å, $b = 10.566(9)$ Å, and $c = 4.390(3)$ Å for b-AsP, respectively. Figure 3 (g) show X-ray diffraction spectrum of b-P bulk single crystal. All the diffraction peaks are quite narrow and can be indexed to b-P crystals with a preferred orientation of (0k0), indicating crystallinity. The Raman spectrum (Fig. 3(f)) shows three distinct vibration modes corresponding to the $A^1_g$, $B_{2g}$, and $A^2_g$ phonon modes of b-P. The XRD and Raman characterization of b-AsP crystals have been reported in our earlier publication [67]. The EDS analysis of b-AsP crystals indicates that the chemical composition is close to b-As$_{0.25}$P$_{0.75}$ (Figure 3(f)). No other additional elements were observed in either b-P or b-AsP crystalline products. Attempts to grow crystals with higher As concentrations were unsuccessful. Indeed, a progressive increase of As content leads to poor crystallinity and the formation of As clusters. It is widely known that the interaction of b-P with H$_2$O and UV light makes it vulnerable to surface oxidation [68]. In addition to modifying the bandgap, the incorporation of As into the b-P lattice greatly increases its surface stability [67].

Our most recent density functional theory calculations for the b-$As_{0.25}P_{0.75}$ monolayer, bilayer, trilayer, and bulk crystals showed the following gaps: 1.558 eV, 1.123 eV, 0.84 eV and 0.21 eV [67]. In general, the band gaps of b-$As_{0.25}P_{0.75}$ are smaller than those of pure b-P. Prototype field-effect transistors made from a few-layers of b-$As_{0.25}P_{0.75}$ exfoliated onto Si/$SiO_2$ substrates exhibit hole-doped like conduction with an ON/OFF current ratio of ~$10^3$ at room temperature and an intrinsic field-effect mobility approaching 300 $cm^2$ $V^{-1}$ $s^{-1}$ at 300 K, which increases up to 600 $cm^2$ $V^{-1}$ $s^{-1}$ at 100 K. These values are comparable or even greater than those reported for pristine b-P. Overall, the anisotropic nature of b-AsP and its ability to alter its band gap make it a promising candidate for fabricating not only transistors and mid-infrared photodetectors, but also as catalysts, non-linear photonic materials, heat conductors, battery, thermoelectric-, and photovoltaic devices [69].

*3.3. High pressure melt growth and characterization of GeAs crystals*

Group IV-V-type 2D materials are a novel family of low-symmetry 2D structures, which have recently garnered growing attention stemming from their outstanding optical and electrical properties [70,71]. Among them, germanium arsenide (GeAs), has emerged as an intriguing layered compound with a crystal structure belonging to the centrosymmetric monoclinic *C*2/*m* (No. 12) space group and exhibiting a considerable in-plane anisotropy [72,73]. In GeAs, every Ge atom is coordinated to three As atoms and another Ge atom, while every As atom is coordinated to three Ge atoms (Figure 4(a)). Each GeAs monolayer is terminated by As atoms and interacts weakly with neighboring layers by van der Waals forces.

As a matter of fact, the GeAs compound melts congruently into a liquid phase at ambient pressure [74]. This allows to grow GeAs crystals under conventional conditions. Nevertheless, the growth from the melt may be challenging due to the presence of GeAs and $GeAs_2$ phases, both congruently melting in the same temperature range, which could result in intergrowing. While the vapor pressure of As is more than two order of magnitude lower than that of P (at the same temperature), the high reactivity and toxicity of As require the use of closed systems to prevent the vapor phase from escaping. In this regard, the use of HPHT technique is highly beneficial.

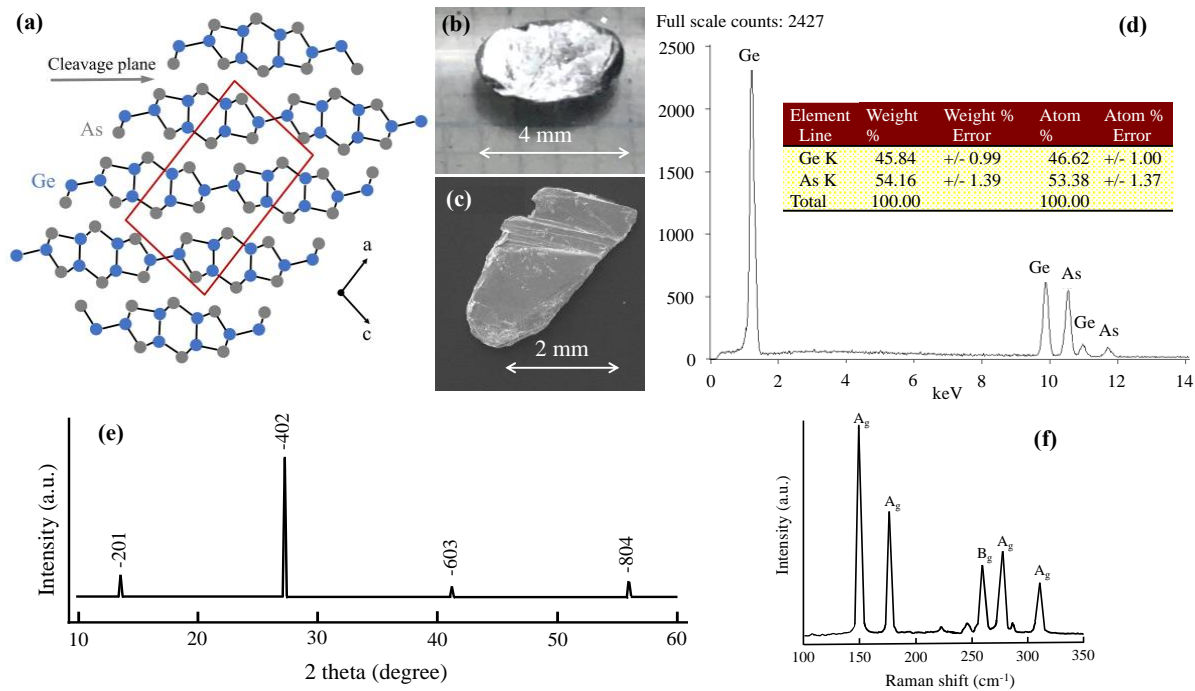

**Figure 4.** (**a**) Crystal structure of GeAs, easy cleavage plane indicated by arrow. (**b**) Optical image of GeAs bulk crystal grown under high pressure. (**c**) Scanning electron microscopy image of GeAs crystal mechanically separated from the bulk crystal and used for EDS analysis. (**d**) EDS analysis of a GeAs single crystal. The material contains only Ge and As elements with the table indicating a composition that is close to $Ge_{0.47}As_{0.53}$. (**e**) X-ray diffraction pattern of GeAs bulk single crystal. (**f**) Raman spectra of GeAs bulk single crystal.

In our standard growth procedure, a high-pressure cell containing the pellet made of Ge and As elements was cold pressurized to 1 GPa, then the temperature was increased to 1000 °C in 1 h, held for 1 h and then slowly (~50 °C/h) cooled to a temperature 650 °C before being quenched to room temperature. The relatively slow cooling rate allows the crystals to nucleate and grow. The obtained GeAs crystals were of good structural quality, have large sizes and shiny surfaces (Figure 4(b,c)). They could be easily cleaved into thin flakes due to the low interlayer energy (~0.19 eV/atom). The products we obtain are similar to those reported in ref. [75], although the size of our crystals is larger, presumably because of the larger size of our BN crucible. The SEM image shown in Figure 4(c), evidences the lamellar structure of the crystals, with large and flat surfaces. EDS analysis, based on the average data collected from five different spots of the crystal, confirms the expected 1:1 ratio while, with a slight increase in As content, the true stoichiometry becomes $Ge_{0.47}As_{0.53}$. The X-ray diffraction pattern obtained from the crystal surface of GeAs bulk crystal are in agreement with previously published results [71] and its narrow reflections are indicative of high crystallinity order. The narrow and well-defined Raman peaks shift further demonstrated the overall high quality of the GeAs crystals produced by HPHT method.

In theory, Ge can form binary compounds with many elements of the IV A group, such as C, Si, Sn, and Pb; of VA group, such as N, P, As, Sb, and Bi; and of VIAgroup, such as O, S, Se, and Te [76]. Like for GeAs, these Ge-based binary compounds may form 2D crystalline materials with various structures and their experimental exploration including by HPHT method looks highly attractive.

**4. Conclusions**

In this study, we demonstrated the usefulness of high-pressure, high-temperature conditions for exploring the crystal growth of 2D van der Waals materials, such as hBN, Mg-hBN, b-P, b-AsP, and GeAs. The hBN and Mg-hBN crystals were produced through exploration of the Mg-B-N solvent system, while b-P, b-AsP, and GeAs were grown from the corresponding elements. All the obtained crystals exhibit a high crystalline quality, are appropriately sized, can be easily exfoliated, and may be used in a variety of measuring techniques to assess how the crystal structure, composition, and thickness of the layers affect the material's physical properties.

The results obtained in the current work can be seen as additional evidence supporting the effectiveness of high-pressure method in the development of 2D materials. Nevertheless, even with these recent advancements, there is still potential for improving the HPHT synthesis of 2D materials. In the case of hBN, one issue is the existence of regions containing carbon and oxygen impurities. These can influence the optical properties of hBN, and it is expected that they also can impact the interfaces with other 2D crystals in van der Waals heterostructures. Even though impurity regions cannot be completely eliminated during synthesis, they can be avoided thereafter by employing cathodoluminescence and photoluminiscence characterization approaches, both developed especially for the rapid selection of crystals or flakes with sizable impurity-free areas. One approach is to include these characterization techniques into an intricate robotic transfer process to improve the selection of hBN flakes [77]. The effective doping of b-P with As opens up the possibility of introducing other substituents into its crystal lattice. This can be used to tune the structural and electrical properties. High pressure may be beneficial for both broadening the choice of substituents and/or for raising the degree of substitution. This is an ongoing topic of study, and there may be new breakthroughs beyond the current state of the art. Overall, the quick progress and the extensive research in this area suggest that 2D materials will play a significant role in shaping the future of science and technology.

**5. Original article statement**

I declare that this manuscript is original, has not been published before and is not currently being considered for publication elsewhere.

**CRediT authorship contribution statement**

N.D. Zhigadlo: Conceptualization, Methodology, Investigation, Data curation, Formal analysis, Visualization, Writing-original draft, Writing - Review & Editing.

**Data availability**

Data will be made available on request.

**Declare of Competing Interest**

The author declare that he has no known competing financial interests or personal relationships that could have appeared to influence the work reported in this paper.

**Acknowledgements**

The author acknowledges support from the Department of Earth Sciences ETH Zurich and the Department of Chemistry and Biochemistry of the University of Bern where these studies were initiated. The author would like to thank T. Shiroka for the critical reading of the manuscript and useful suggestions.